\documentclass[aps,pra,twocolumn,superscriptaddress,showpacs,amsmath]{revtex4}		 
\usepackage{graphicx}
 
\begin{document}

\title{Phase-dependent fluctuations of intermittent resonance fluorescence}  

\author{H\'ector M. Castro-Beltr\'an} 		 
\email{hcastro@uaem.mx} 
\affiliation{Centro de Investigaci\'on en Ingenier\'{\i}a y Ciencias Aplicadas, \\
Instituto de Investigaci\'on en Ciencias B\'asicas y Aplicadas, \\
Universidad Aut\'onoma del Estado de Morelos, 
Avenida Universidad 1001, 62209 Cuernavaca, Morelos, M\'exico}
\affiliation{Instituto de Ciencias F\'isicas, 
Universidad Nacional Aut\'onoma de M\'exico, \\
Apartado Postal 48-3, 62251 Cuernavaca, Morelos, M\'exico}
\author{Ricardo Rom\'an-Ancheyta} 
\email{ancheyta6@gmail.com} 
\affiliation{Instituto de Ciencias F\'isicas, 
Universidad Nacional Aut\'onoma de M\'exico, \\
Apartado Postal 48-3, 62251 Cuernavaca, Morelos, M\'exico}
\author{Luis Guti\'errez}  
\email{luis.gutierrez@uaem.mx}
\affiliation{Centro de Investigaci\'on en Ingenier\'{\i}a y Ciencias Aplicadas, \\
Instituto de Investigaci\'on en Ciencias B\'asicas y Aplicadas, \\
Universidad Aut\'onoma del Estado de Morelos, 
Avenida Universidad 1001, 62209 Cuernavaca, Morelos, M\'exico}

\date{\today}
\begin{abstract} 
Electron shelving gives rise to bright and dark periods in the resonance 
fluorescence of a three-level atom. The spectral signature of such blinking 
is a very narrow inelastic peak on top of the two-level atom spectrum. Here, 
we investigate theoretically phase-dependent fluctuations (e.g., squeezing) 
of intermittent resonance fluorescence in the frameworks of balanced and 
conditional homodyne detection (BHD and CHD, respectively). In BHD, the 
squeezing is reduced significantly in size and Rabi frequency range 
compared to that for a two-level atom. The sharp peak is found only in the 
spectrum of the squeezed quadrature, splitting the negative broader 
squeezing peak for weak fields. CHD correlates the BHD signal with the  
detection of emitted photons. It is thus sensitive to third-order fluctuations 
of the field, produced by the atom-laser nonlinearity, that cause noticeable 
deviations from the second-order BHD results. For weak driving, the 
third-order spectrum is negative, enlarging the squeezing peak but also 
reducing the sharp peak. For strong driving, the spectrum is dominated by 
third-order fluctuations, with a large sharp peak and the sidebands becoming 
dispersive. Finally, the addition of third-order fluctuations makes the 
integrated spectra of both quadratures equal in magnitude in CHD, in 
contrast to those by BHD.  A simple mathematical approach allows us to 
obtain very accurate analytical results in the shelving regime. 
\end{abstract}

\maketitle

\section{\label{sec:intro}Introduction}  
A photon emitter with peculiar fluctuations is a single three-level atom with 
a laser-driven strong transition competing with a coherently or incoherently  
driven weak transition. The occasional population of a long-lived state, an 
effect called electron shelving, produces intermittence (blinking) in the 
resonance fluorescence of the strong transition. Photon statistics of the 
fluorescence have been thoroughly studied for three-level atomic systems 
\cite{PlKn97}, in which case the process is ergodic, i.e., when the mean 
bright and dark periods are finite. For a single quantum dot or molecule the 
statistics are more complicated if the process is not ergodic \cite{StHB09}.  
In the spectral domain, ergodic shelving manifests in the appearance of a 
very narrow inelastic peak on top of the central peak of the two-level-like 
spectrum. This has been well studied analytically and numerically 
\cite{HePl95,GaKK95,EvKe02} and observed experimentally \cite{BuTa00}. 
In the latter, heterodyne detection was used, which allows for very high 
spectral resolution \cite{HBLW97}. In their paper \cite{BuTa00}, B\"uhner 
and Tamm suggest performing complementary phase-dependent 
measurements of the fluorescence; so far, there are no reports yet, 
perhaps due to experimental restrictions.  

Squeezing, the reduction of fluctuations below those of a coherent state 
in a quadrature at the expense of increasing fluctuations in the other 
quadrature, is weak in resonance fluorescence \cite{WaZo91,CoWZ84}. 
The low collection and imperfect quantum efficiency of photodetectors 
have been the main barriers for the observation of squeezing, although 
recent experimental progress tackle these issues. On the one hand, 
there is the increased solid angle of emission captured with minimal 
disturbance of the photon density of states surrounding the atom 
\cite{SoLe15}. On the other hand, there is the development of conditional 
detection schemes based on homodyne detection that cancel the finite 
quantum efficiency issue 
\cite{Vogel91,Vogel95,CCFO00,FOCC00,HSL+06,GSS+07,KAD+09,KuVo-X}. 
We discuss two of them. 

Homodyne correlation measurement (HCM), proposed by Vogel 
\cite{Vogel91,Vogel95} (see also \cite{Carm85}), consists of intensity 
correlations of the previously mixed source and weak local oscillator  
fields, thus canceling the detector efficiency factors. The output contains 
several terms, including the variance and an amplitude-intensity correlation. 
Very recently, HCM was used to observe squeezing in the resonance 
fluorescence of a single two-level quantum dot \cite{SHJ+15} in conditions 
close to those for free-space atomic resonance fluorescence. In fact, in 
the first demonstration of HCM the amplitude-intensity correlation of the 
fluorescence of a single three-level ion in the $\Lambda$ configuration 
was observed \cite{GRS+09} although not yet in the squeezing regime. 

Conditional homodyne detection (CHD) was proposed and demonstrated 
by Carmichael, Orozco and coworkers \cite{CCFO00,FOCC00}. This 
consists of balanced homodyne detection (BHD) of a quadrature 
conditioned on an intensity measurement of part of the emitted field; 
it gives the amplitude-intensity correlation of HCM but measured directly, 
without the other terms. As in the intensity correlations, the conditioning 
cancels the dependence on detector efficiency. The intensity detection 
channel has nontrivial effects on the quadrature signals. The 
amplitude-intensity correlation is of third order in the field amplitude; 
hence it allows for third-order fluctuations. Initially, CHD was devised for 
weak light emitters, neglecting the third-order fluctuations. This allowed 
the identification of the Fourier transform of the correlation as the spectrum 
of squeezing \cite{CCFO00,FOCC00}. However, recent work on CHD of 
two-level atom resonance fluorescence has shown important deviations 
from the spectrum of squeezing due to increasing nonlinearity in the 
atom-laser interaction \cite{hmcb10,CaGH15}. An additional display of 
these non-Gaussian fluctuations is found in the asymmetry of the 
correlation in cavity QED \cite{DeCC02} and in the resonance 
fluorescence of a $V$-type three-level atom 
\cite{MaCa08,CaGM14,XGJM15} and of two blockading Rydberg atoms 
\cite{XuMo15}.  

In this paper we investigate theoretically ensemble-averaged 
phase-dependent fluctuations of the intermittent (ergodic) resonance 
fluorescence of a single three-level atom (3LA). Besides numerical solutions 
for the one- and two-time expectation values, we obtain approximate 
analytical solutions which are very accurate in the limit when the decay rate 
of the strong transition is much larger than those of the weak transitions. 
Our solutions are simple and reflect clearly the time and spectral scales. 
Thus we begin by writing the expression for the coherent and 
phase-independent incoherent spectra of the 3LA, studied numerically at 
length in Ref. \cite{EvKe02}.   

We compare the spectra and variances of an ideal BHD approach, which 
could also be obtained from HCM, with those of the CHD method. They 
have in common that the sharp extra peak 
\cite{HePl95,GaKK95,EvKe02,BuTa00}, on atom-laser resonance, is a 
feature \textit{only} of the quadrature that features squeezing; in the other 
quadrature the spectrum is a simple broad positive Lorentzian. In the 
weak-field limit, while both methods give similar negative spectra for a 
two-level atom (2LA), the sharp peak is positive, reducing the squeezing in 
BHD and enhancing the negative peak in CHD. For a strong laser field the 
third-order fluctuations of CHD distort the positive Lorentzian sidebands of 
the Mollow triplet and turn them dispersive for both 2LA and 3LA. However, 
for the 3LA, both the sharp peak and the dispersive sidebands are much 
larger than the second-order spectrum. 

Interestingly, in CHD, the addition of third-order fluctuation makes the 
integrated spectra of both quadratures equal in magnitude, in contrast to 
the case of the spectrum by BHD \cite{CaGH15}. This feature of CHD may 
be a bonus over other modern variations of the standard homodyne 
detection scheme.   

This paper is organized as follows: In Sec. II we introduce the atom-laser 
model and obtain approximate analytic solutions in the shelving regime. 
In Sec. III we calculate the phase-independent spectrum, and in Sec. IV 
we calculate the phase-dependent spectra and variances. Sections V and 
VI are devoted to the amplitude-intensity correlation by CHD and its 
spectrum, respectively. Finally, conclusions are given in Sec. VII. Two 
appendices summarize the analytic and numerical methods employed.

\section{\label{sec:model}Atom-Laser Model and Solutions}
\begin{figure}[t]
\includegraphics{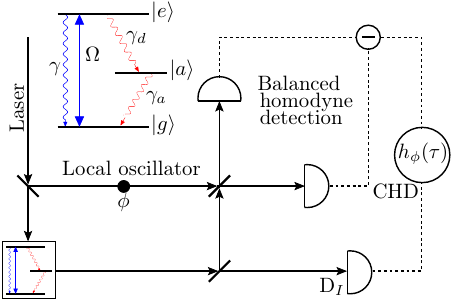} 
\caption{\label{fig:chdsetup} 
(Color online) Scheme of conditional homodyne detection. Blocking the 
path to the lower detector, $\mathrm{D}_I$, realizes the standard balanced 
homodyne detection. The inset shows the three-level atom-laser interaction 
and spontaneous decays. }
\end{figure}
We consider a single three-level atom where a laser of Rabi frequency 
$\Omega$ drives a transition between the ground state $|g \rangle$ and an 
excited state $|e \rangle$. The excited state has two spontaneous emission 
channels: one directly to the ground state with rate $\gamma$ for the driven 
transition, and one via a long-lived shelving state $|a \rangle$ with rate 
$\gamma_d$, which in turn decays to the ground state with rate 
$\gamma_a$ (see Fig. \ref{fig:chdsetup}). In the limit 
\begin{eqnarray}    \label{eq:gammas}
\gamma \gg \gamma_d \,, \gamma_a		  	
\end{eqnarray} 
the fluorescence of the driven transition features well-defined bright and 
dark periods of average lengths,
\begin{eqnarray}    	\label{brightdarktimes}
T_B = \frac{2\Omega^2 +\gamma^2}{\gamma_d \Omega^2} \,, 
\qquad  		T_D &=& \gamma_a^{-1} \,, 
\end{eqnarray}
respectively, as calculated in Ref. \cite{EvKe02} using a random telegraph 
model.   

Throughout this paper we assume zero atom-laser detuning. This serves 
two purposes: first, we limit the discussion to the essentials of the main 
topics; second, with further assumptions discussed later, we obtain close 
approximate analytical solutions. The master equation for the atomic 
density operator, in the frame rotating at the laser frequency, can be 
written as   
\begin{eqnarray} 	\label{masterEq}
\dot{\rho}(t) &=& 
  -i \frac{\Omega}{2} [\sigma_{eg} +\sigma_{ge} ,\rho]    	\nonumber \\ 
&& +\frac{\gamma}{2} \left( 2\sigma_{ge} \rho \sigma_{eg} 
  -\sigma_{ee} \rho -\rho \sigma_{ee} \right) 		\nonumber \\ 
&& + \frac{\gamma_d}{2} \left( 2\sigma_{ae} \rho \sigma_{ea} 
  -\sigma_{ee} \rho -\rho \sigma_{ee} \right) 		\nonumber \\ 
&& + \frac{\gamma_a}{2} \left( 2\sigma_{ga} \rho \sigma_{ag} 
  -\sigma_{aa} \rho -\rho \sigma_{aa} \right) \,, 
\end{eqnarray}
where $\sigma_{jk} = | j \rangle \langle k |$ are atomic transition operators 
which obey the inner product prescription $\langle j | k \rangle =\delta_{jk}$. 

We obtain two sets of equations. The first one is  
\begin{subequations} 
\begin{eqnarray} 	\label{eq:BlochEqs} 
\dot{\rho} &=& \mathbf{M}  {\rho} +\mathbf{b} 	\,, 	
\end{eqnarray} 
where $\rho = ( \rho_{eg}, \rho_{ge}, \rho_{ee}, \rho_{gg} )^T$, 
$\mathbf{b} = ( 0, 0, 0, \gamma_a )^T$,  and 
\begin{eqnarray} 	\label{eq:matrixM} 
\mathbf{M} &=& \left( \begin{array}{cccc} 
	-\gamma_+/2 & 0 & i\Omega/2 & -i\Omega/2 \\
	0 & -\gamma_+/2 &  -i\Omega/2 & i\Omega/2 \\ 
	i\Omega/2 & -i\Omega/2 & -\gamma_+ & 0 \\ 
	-i\Omega/2 & i\Omega/2 & \gamma_- & -\gamma_a 
	\end{array} 	\right) 	\,,
\end{eqnarray} 
\end{subequations}
where
\begin{eqnarray}    	
\gamma_+ = \gamma +\gamma_d 	\,, 	\qquad 
\gamma_- = \gamma -\gamma_a 	\,.   
\end{eqnarray} 
Here, we have eliminated the population $\rho_{aa}$ due to 
conservation of probability, $\rho_{gg} +\rho_{ee} +\rho_{aa} =1$. 

The second set of equations involves the coherences linking states 
$|e \rangle$ and $|g \rangle$ to state $|a \rangle$, i.e.,   
$(\rho_{ga}, \rho_{ag},\rho_{ea},\rho_{ae})^T$. They evolve with damped 
oscillations with zero mean. The two sets are decoupled, and only the 
first one is relevant for the purposes of this work.

We obtain first the steady state of the density operator (labeled with the 
abbreviation $st$). For a more compact notation we define  
$\alpha_- =\rho_{eg}^{st} =\langle \sigma_- \rangle_{st}$, 
$\alpha_+ =\alpha_-^{\ast}$, and 
$\alpha_{jj} =\rho_{jj}^{st} =\langle \sigma_{jj} \rangle_{st}$.  
We have   
\begin{subequations} 	\label{eq:alphas}
\begin{eqnarray} 	
\alpha_{\mp} &=& \mp i \frac{ Y/\sqrt{2}}{ 1+Y^2 +(q/2)Y^2 } \,, 
	\label{eq:alphapm} \\ 
\alpha_{ee} &=& \frac{Y^2/2}{1+Y^2 +(q/2)Y^2 } \,, 	\label{eq:alphaee} \\ 
\alpha_{gg} &=& \frac{ 1+ Y^2/2 }{1+Y^2 +(q/2)Y^2}  \,, \label{eq:alphagg}	\\
\alpha_{aa} &=& q \alpha_{ee} 	\,, 	\label{eq:alpha_aa}
\end{eqnarray}
\end{subequations}
where 
\begin{eqnarray}    	
q=\gamma_d/\gamma_a \,,   		\qquad 
Y = \sqrt{2} \Omega/\gamma_+ 	\,.  	
\end{eqnarray} 
For $\gamma_d=0$ ($q=0$) we recover the results of the 2LA. 

Equation (\ref{eq:BlochEqs}) is still too complicated to solve  analytically 
in the general case. However, in the limit (\ref{eq:gammas}), very good 
approximate solutions are obtained (see Appendix \ref{sec:approx} for 
more details). We use a Laplace transform approach to obtain 
approximate expectation values of the atomic vector, 
$\mathbf{s} = (\sigma_-, \sigma_+, \sigma_{ee}, \sigma_{gg} )^T$, 
and two-time correlations.  With the atom initially in its ground state, 
$\langle \mathbf{s}(0) \rangle= ( 0,0,0,1 )^T$, the expectation values 
of the atomic operators are 
\begin{subequations} 	\label{eq:BlochSols}
\begin{eqnarray} 	
\langle \sigma_{\mp} (t) \rangle 
&=& 	\mp i \frac{ Y/ \sqrt{2} }{1+Y^2}  f(t)  
\mp i  \frac{ \sqrt{2} \gamma_+ Y }{8 \delta} 
	\left( e^{\lambda_+ t} -e^{\lambda_- t} \right)	\nonumber \\ 
&& 	+\alpha_{\mp} \left( 1- e^{\lambda_2 t} \right) \,,  \\   
\langle \sigma_{ee} (t) \rangle &=& 	
   \frac{ Y^2/2 }{1+ Y^2} f(t) +\alpha_{ee} \left( 1- e^{\lambda_2 t} \right) \,, 	\\ 
\langle \sigma_{gg} (t) \rangle 
	&=& e^{\lambda_2 t}  -\frac{ Y^2/2 }{1+ Y^2} f(t)    
  	+\alpha_{gg} \left( 1- e^{\lambda_2 t} \right) \,, 	 
\end{eqnarray}
\end{subequations} 
where 
\begin{eqnarray} 	
f(t) &=& e^{\lambda_2 t} -\frac{1}{2}  \left[ 
	\left( 1+\frac{3 \gamma_+}{4 \delta} \right) e^{\lambda_+ t}  
	\right.	\nonumber \\ 		
&& \left. + \left( 1-\frac{3 \gamma_+}{4 \delta} \right) e^{\lambda_- t}  
	\right]   \,, 
\end{eqnarray}
\begin{subequations}  	\label{eq:eigenvalues}
\begin{eqnarray} 	
\lambda_1 &=& -\gamma_+/2 \,, 	\\ 
\lambda_2 &=& -\gamma_a \left( 1+ q \frac{\Omega^2 }
	{ 2 \Omega^2 +\gamma^2 }  \right)  \,, \label{eq:ev2}	\\  
\lambda_{\pm} &=& -\frac{3\gamma_+}{4} \pm \delta  \,,  
\end{eqnarray} 
\end{subequations} 
and 
\begin{eqnarray} 	
\delta &=& (\gamma_+/4) \sqrt{1-8Y^2} 	\,. 
\end{eqnarray}

This approach allows us to identify Eqs. (\ref{eq:eigenvalues}) as the 
eigenvalues of the matrix (\ref{eq:matrixM}) of the master equation. This is 
much more convenient than attempting to write the exact ones in compact 
form. The eigenvalues contain the kernel of the atomic evolution, that is, 
the scales of decay and coherent evolution, as well as the corresponding 
widths and positions of the spectral components. 

The first eigenvalue, $\lambda_1$, is exact and gives half the total decay rate 
from the excited state. Although absent in Eqs. (\ref{eq:BlochSols}), it occurs in 
the second-order correlations (see below). Then, $\lambda_2$ represents the 
slow decay rate due to shelving. This caues the steady state to be reached 
after a long time, $t \sim \gamma_d^{-1}$. Borrowing from the random 
telegraph model \cite{EvKe02}, the slow decay rate is given by 
$\lambda_2 = -(T_D^{-1} +T_B^{-1})$. The two remaining eigenvalues 
represent the damped coherent evolution;  they are real if $8Y^2 \le 1$ and 
complex if $8Y^2 >1$. Eigenvalues $\lambda_1, \lambda_{\pm}$ contain the 
two-level-like evolution towards a quasi-steady state (with the decay rate 
$\gamma$ of the two-level case replaced by $\gamma_+$ for the 3LA) that 
is followed by the slow decay. 

The two-time correlations 
$\langle \sigma_+(0) \mathbf{s} (\tau) \sigma_-(0) \rangle_{st}$, which 
have initial conditions $(0,0,0,\alpha_{ee} )^T$, are approached like those 
for $\langle \mathbf{s} (t) \rangle$. Using the quantum regression formula 
(see, e.g., \cite{Carm99}) and $\mathbf{s}(0)= (0,0,0,1)^T$, we have 
\begin{eqnarray} 	\label{eq:3rdordercorr}
\langle \sigma_+(0) \mathbf{s} (\tau) \sigma_-(0) \rangle_{st} 
	= \alpha_{ee}  \langle  \mathbf{s}(\tau) \rangle_{ \mathbf{s}(0)  } 	\,,    
\end{eqnarray}
that is, these correlations are identical to Eqs. (\ref{eq:BlochSols}) times the  
factor $\alpha_{ee}$, with $t$ replaced by $\tau$. 

The approximate analytic solutions to the correlations 
$\langle \sigma_+(0) \mathbf{s} (\tau) \rangle_{st}$, which have initial 
conditions $(\alpha_{ee},0,0, \alpha_+)^T$, can be similarly obtained (see 
Appendix \ref{sec:approx}). We use them, however, to obtain the solutions 
for correlations of fluctuations, 
$\langle \Delta \sigma_+(0) \Delta \mathbf{s} (\tau) \rangle_{st}$, where   
\begin{eqnarray} 	\label{eq:dipoleSplit}
\Delta \sigma_{jk} (t) = \sigma_{jk} (t)  -\langle \sigma_{jk}  \rangle_{st} \,, 	
	\qquad \langle \Delta \sigma_{jk}(t)\rangle =0 	\,. 
\end{eqnarray}
Hence    
\begin{eqnarray} 		\label{eq:corr2split}
\langle \Delta \sigma_+(0) \Delta \sigma_{\mp}(\tau) \rangle_{st} 
	&=& \langle \sigma_+(0) \sigma_{\mp}(\tau) \rangle_{st}  
	-\langle \sigma_+ \rangle_{st}  \langle \sigma_{\mp} \rangle_{st} \,, 
	\nonumber \\ 
\end{eqnarray}
yielding  
\begin{eqnarray} 	\label{eq:p_tau}	
\langle \Delta \sigma_+(0) \Delta \sigma_{\mp}(\tau) \rangle_{st} 
&=& C_1 e^{\lambda_1 \tau} \pm C_2 e^{\lambda_2 \tau} 
		\nonumber \\ 
&&	\mp C_+ e^{\lambda_+ \tau}  \mp  C_- e^{\lambda_- \tau} 	\,, 
\end{eqnarray}
where 
\begin{subequations} 	\label{eq:p-coef}
\begin{eqnarray} 		
C_1  &=& \frac{Y^2/4}{ 1+Y^2 +(q/2)Y^2 } 	\,, \\ 
C_2 &=&  \frac{q Y^4/4}{ (1+Y^2) \left( 1+Y^2 +(q/2)Y^2 \right)^2}	\,,  
	\label{eq:C2} \\ 
C_{\mp} &=&   \frac{ Y^2[1-Y^2 \pm (1-5Y^2)(\gamma_+/4\delta)] }
	{ 8(1+Y^2) \left( 1+Y^2 +(q/2)Y^2 \right)} 	  \,.
\end{eqnarray} 
\end{subequations} 

\section{\label{sec:powerspec}Stationary Power Spectrum} 
The stationary (Wiener-Khintchine) power spectrum is given by the Fourier 
transform of the dipole field auto-correlation function,  
\begin{eqnarray}
S(\omega) &=& \frac{1}{\pi \alpha_{ee}} \mathrm{Re} 
	\int_0^{\infty} d\tau e^{-i \omega \tau} 
	\langle \sigma_+ (0) \sigma_- (\tau) \rangle_{st}  	\,.
\end{eqnarray}
The factor $(\pi \alpha_{ee})^{-1}$ normalizes the integral of $S(\omega)$ 
over all frequencies to unity. Equation (\ref{eq:corr2split}) separates the 
spectrum in two parts: 
\begin{eqnarray}
S(\omega) &=& S_{coh}(\omega) +S_{inc}(\omega) 	\,,
\end{eqnarray}
where 
\begin{eqnarray} 	\label{eq:ScohDef}
S_{coh}(\omega) &=& \frac{|\alpha_{+}|^2}{\pi \alpha_{ee}} \mathrm{Re} 
	\int_0^{\infty}  e^{-i \omega \tau} d\tau 	
= \frac{|\alpha_{+}|^2}{\pi \alpha_{ee}}  \delta(\omega)	 		
\end{eqnarray}
and 
\begin{eqnarray} 	\label{eq:SincDef}
S_{inc}(\omega) &=& 
\frac{1}{\pi \alpha_{ee}} \mathrm{Re} \int_0^{\infty}  d\tau e^{-i \omega \tau} 
	\langle \Delta \sigma_+(0) \Delta \sigma_-(\tau) \rangle_{st}  	
	\nonumber 	 \\
\end{eqnarray}
are, respectively, the coherent spectrum, due to elastic scattering, and 
the incoherent (inelastic) spectrum, due to atomic fluctuations.  

\begin{figure}[t]
\includegraphics[width=8.5cm]{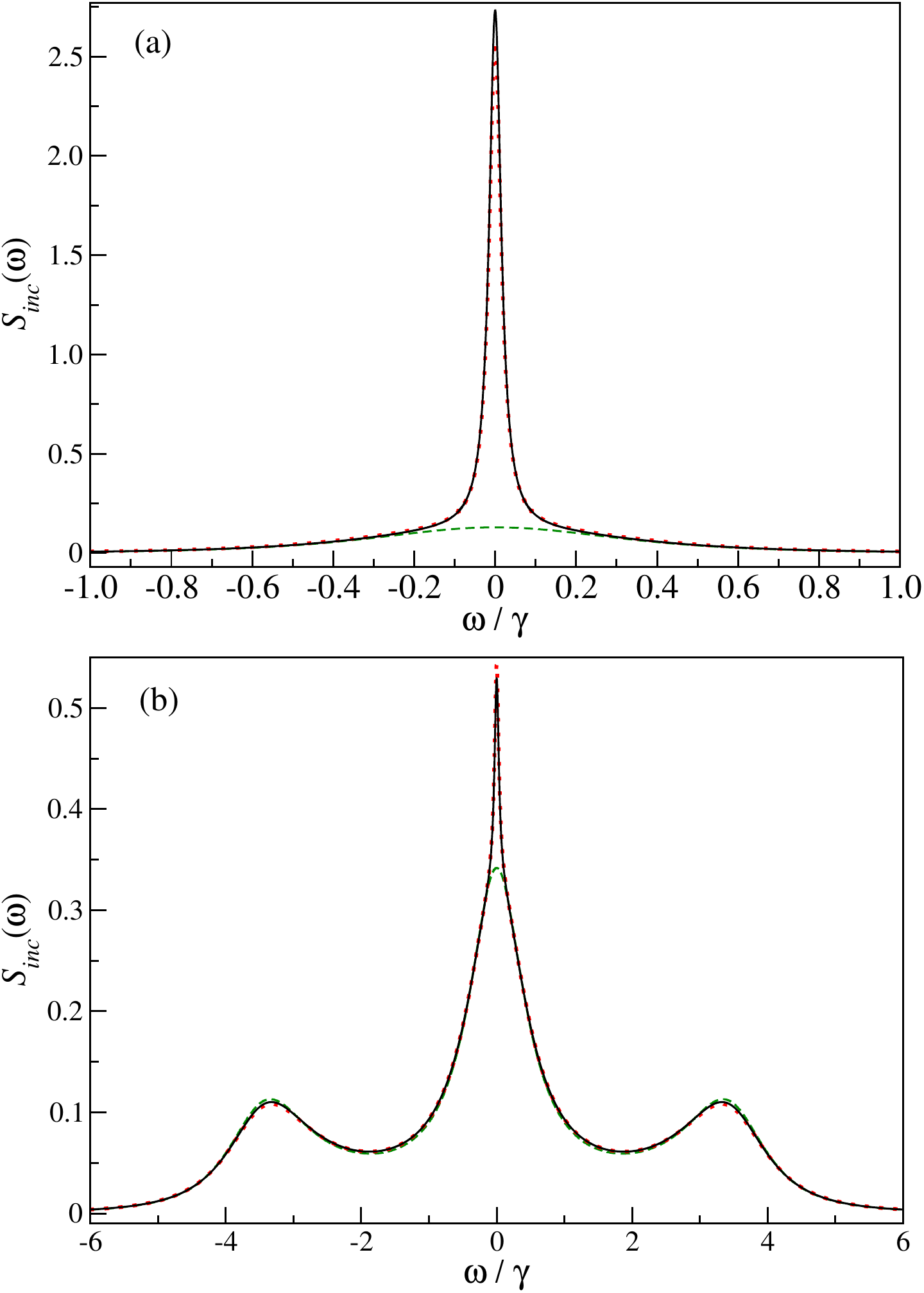} 
\caption{\label{fig:Sinc} 
(Color online) Incoherent spectrum for (a) a saturating laser field, 
$\Omega =\gamma_+/4 =0.2625 \gamma$, and (b) a strong field, 
$\Omega=3.5 \gamma$, with $\gamma_d=0.05 \gamma$ and 
$\gamma_a=0.015 \gamma$. The solid black and dotted red curves 
are, respectively, the exact and approximate spectra, and the dashed 
green curve is the 2LA spectrum. }
\end{figure}
The main features of the spectrum of the atom-laser system of the previous 
section were studied in \cite{EvKe02}. The incoherent spectrum consists 
of a two-level-like structure that becomes a triplet for strong excitation 
\cite{Mollow69}, plus a sharp peak, associated with the eigenvalue 
$\lambda_2$, due to the shelving of ethe lectronic population in the 
long-lived state. This three-level system contains the essential physics of 
the more complex atomic system used for the experimental observation 
of the sharp peak \cite{BuTa00} by heterodyne detection, able to resolve 
hertz or sub-hertz features \cite{HBLW97}. The sharp peak had been 
predicted for the $V$-type and $\Lambda$-type 3LAs \cite{HePl95,GaKK95}, 
which also feature electron shelving. 

Our Laplace transform approach allowed us to obtain a very good analytic 
approximation to the full spectrum, split into its various components, with 
their widths and amplitudes readily spotted. Substituting Eq. (\ref{eq:p_tau})   
into Eq. (\ref{eq:SincDef}) the incoherent spectrum is  
\begin{eqnarray} 	\label{eq:3LAincSpec}
 S_{inc}(\omega) &=& \frac{1}{\pi \alpha_{ee}} \left[ 
	C_+ \frac{ \lambda_+}{\omega^2 +\lambda_+^2}  
	+C_- \frac{ \lambda_-}{\omega^2 +\lambda_-^2}  \right. \nonumber \\ 
&& \left. -C_1 \frac{ \lambda_1}{\omega^2 +\lambda_1^2} 
	-C_2 \frac{ \lambda_2}{\omega^2 +\lambda_2^2}  \right]	\,.
\end{eqnarray}
In Fig. \ref{fig:Sinc} we plot this spectrum with eigenvalues 
(\ref{eq:eigenvalues}) along the exact and 2LA spectra. It reproduces 
remarkably well the exact spectrum, with the sharp peak being slightly 
smaller (bigger) in the saturating (strong) case than the exact one. Also, 
making $\gamma_d=0$, the formula is exact for the 2LA spectrum 
\cite{Mollow69}. The intensity (integral over all frequencies) of the sharp 
peak is 
\begin{eqnarray*} 	
I_{ep} &=&  \frac{q Y^2/2}{ (1+Y^2) \left( 1+Y^2 +(q/2)Y^2 \right)}	\,. 
\end{eqnarray*}
It is small for both weak and strong driving (proportional to $Y^2$ and
$Y^{-2}$, respectively), and largest for $\Omega \approx 3\gamma/4$. 
 
The coherent spectrum of the 3LA is 
\begin{eqnarray}
S_{coh}(\omega) 
	&=& \frac{1}{\pi \left( 1+Y^2 +(q/2)Y^2 \right)} \delta(\omega) \,, 	
\end{eqnarray}
smaller than that of the 2LA (where $q=0$) \cite{EvKe02}. The difference 
in intensity is precisely given by $I_{ep}$.  

The choice of values $\gamma_d  =0.05 \gamma$ and 
$\gamma_a  =0.015 \gamma$, small enough to fulfill the limit  
(\ref{eq:gammas}), is such that the relation $\gamma_d =3.3 \gamma_a$ 
closely optimizes the intensity of the sharp extra peak for any given Rabi 
frequency \cite{EvKe02}. For simplicity, we use these values for all the 
remaining 3LA plots in this work.

\begin{widetext}
\begin{figure}[h]
\includegraphics[width=17.5cm]{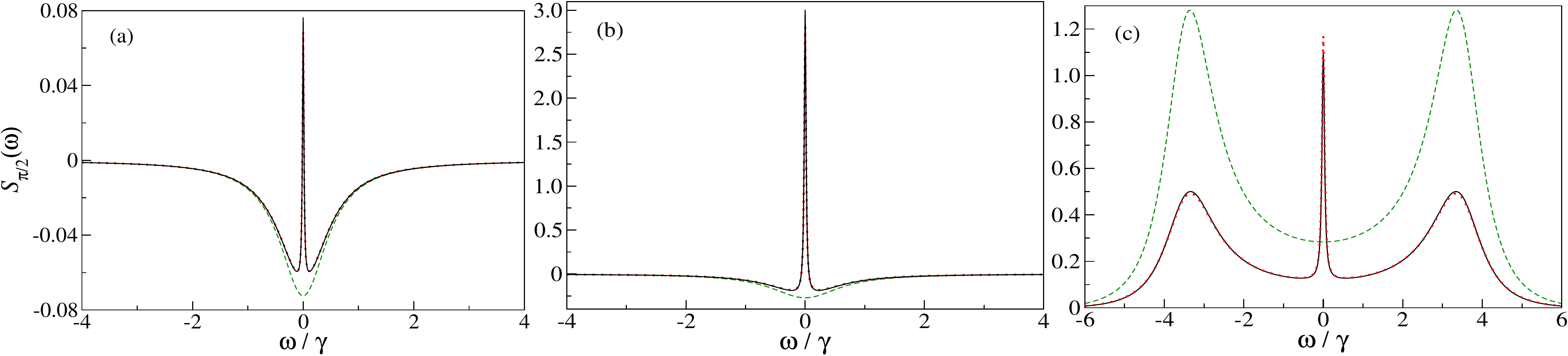}%
\caption{\label{fig:SqueezSpec} 
(Color online) Spectra of the $\phi=\pi/2$ quadrature for 
(a) $\Omega=0.1 \gamma$, (b) $\Omega=0.2625 \gamma$, and 
(c) $\Omega=3.5 \gamma$.  The other parameters are 
$\gamma_d=0.05 \gamma$, $\gamma_a=0.015 \gamma$, and $\eta=1$. 
The solid black and dotted red lines correspond to the exact and 
approximate 3LA spectra, respectively, and the dashed green lines are 
the 2LA spectra.}
\end{figure}
\end{widetext}
\section{\label{sec:specsqueez}The Spectrum of Squeezing} 
Now we turn to the phase-dependent spectrum of the fluorescence of the 
three-level atom and compare it to the well-known case of the two-level 
atom \cite{CoWZ84,RiCa88}. Following Carmichael \cite{Carm87}, we 
define the \textit{ideal source field} spectrum of squeezing as the Fourier 
transform of photocurrent fluctuations of the quadratures in homodyne 
detection, 
\begin{eqnarray} 	\label{eq:specsqueez}
S_{\phi}(\omega) &=& 
	8\gamma_+ \eta \int_{0}^{\infty}  d\tau \cos{\omega \tau} \,	
	 \langle : \Delta \sigma_{\phi}(0) \Delta \sigma_{\phi}(\tau) : \rangle_{st}   
	 \nonumber \\
&=& 8\gamma_+ \eta \int_{0}^{\infty}  d\tau \cos{\omega \tau} 	\nonumber \\ 
	&& \times	\mathrm{Re} \left[ e^{-i\phi}  
	\langle \Delta \sigma_+(0) \Delta \sigma_{\phi}(\tau) \rangle_{st}  \right]  \,, 
\end{eqnarray}
where 
\begin{eqnarray} 	\label{eq:fluctop}
\Delta \sigma_{\phi} &=& \frac{1}{2} \left(\Delta \sigma_- e^{i\phi} 
	+\Delta \sigma_+ e^{-i\phi} \right) \,,  
\end{eqnarray}
$\phi$ is the phase of the local oscillator in a BHD setup (that is, blocking 
the path to detector $\mathrm{D}_I$ in Fig. \ref{fig:chdsetup}), $\eta$ is a 
combined collection and detection efficiency, and the dots $::$ indicate that 
the operators must follow time and normal orderings. This is an 
incoherent spectrum as it depends on the field fluctuations. In fact, the 
phase-dependent and the phase-independent spectra are related as 
\cite{RiCa88}   
\begin{eqnarray} 	 \label{eq:S_RC} 
S_{inc}(\omega) &=& \frac{1}{8\pi \alpha_{ee} \gamma_+ \eta} 
	\left[ S_{\phi}(\omega) +S_{\phi+\pi/2}(\omega)   \right] 	 \,. 	
\end{eqnarray}
Adding the spectra for $\phi=0$ and $\pi/2$ Eq. (\ref{eq:SincDef}) is 
recovered.

Although the atom and laser parameters do not always allow for squeezing 
(negative values in the spectrum), we keep the moniker of spectrum of 
squeezing in order to distinguish this from the spectrum of 
Sec. \ref{sec:chd-spec}. 

Substituting Eq. (\ref{eq:p_tau}) in Eq. (\ref{eq:specsqueez}) the 
approximate spectra for the quadratures are   
\begin{eqnarray} 	 
S_{0}(\omega) &=& -8 \gamma_+ \eta 
	C_1 \frac{ \lambda_1}{\omega^2 +\lambda_1^2} 	\,, 	
	\label{eq:SqSpec_0}  \\
S_{\pi/2}(\omega) &=& 8 \gamma_+ \eta  \left[ 
	C_+ \frac{ \lambda_+}{\omega^2 +\lambda_+^2} 
	+C_- \frac{ \lambda_-}{\omega^2 +\lambda_-^2} 	\right.	
	\nonumber \\ 
&& \left.  -C_2 \frac{ \lambda_2}{\omega^2 +\lambda_2^2}   \right]   \,. 	
	\label{eq:SqSpec_hpi}
\end{eqnarray}

For $\phi=0$ the spectrum is only a single, positive (no squeezing) 
Lorentzian, just like for the 2LA, now with a width of $\gamma_+/2$. 
For $\phi=\pi/2$ the spectrum is more interesting, as shown in 
Fig. \ref{fig:SqueezSpec} for several field strengths. For instance, it has 
a sharp peak [last term in Eq. (\ref{eq:SqSpec_hpi})], with its maximum 
near $\Omega \approx 0.9 \gamma$. From weak to little more than 
saturating fields the first two terms of Eq. (\ref{eq:SqSpec_hpi}) (with 
factors $C_{\pm} |\lambda_{\pm}|$) add to form a single negative peak, 
indicating squeezing. Rice and Carmichael \cite{RiCa88} found that the 
weak-field spectrum ($Y^2 \ll 1$) in the 2LA has a line-width smaller than 
$\gamma/2$ due to the negative value of the Lorentzians with amplitudes 
$C_{\pm} |\lambda_{\pm}|$ in Eq. (\ref{eq:3LAincSpec}), resulting in a 
squared Lorentzian \cite{Mollow69}. In the 3LA there is less squeezing 
and the sharp peak splits the squeezing peak. For strong fields, the 
spectrum consists of the sidebands of the Mollow triplet plus the extra 
peak.  

An additional manifestation of shelving is the shrinking of the sidebands 
of the quadrature spectra compared to those of the 2LA. This is because 
state $| a \rangle$ takes up an important fraction of the steady-state 
population (actually, $\alpha_{aa} = q \alpha_{ee}$) for increasing Rabi 
frequency. 

\begin{figure}
\includegraphics[width=8.5cm]{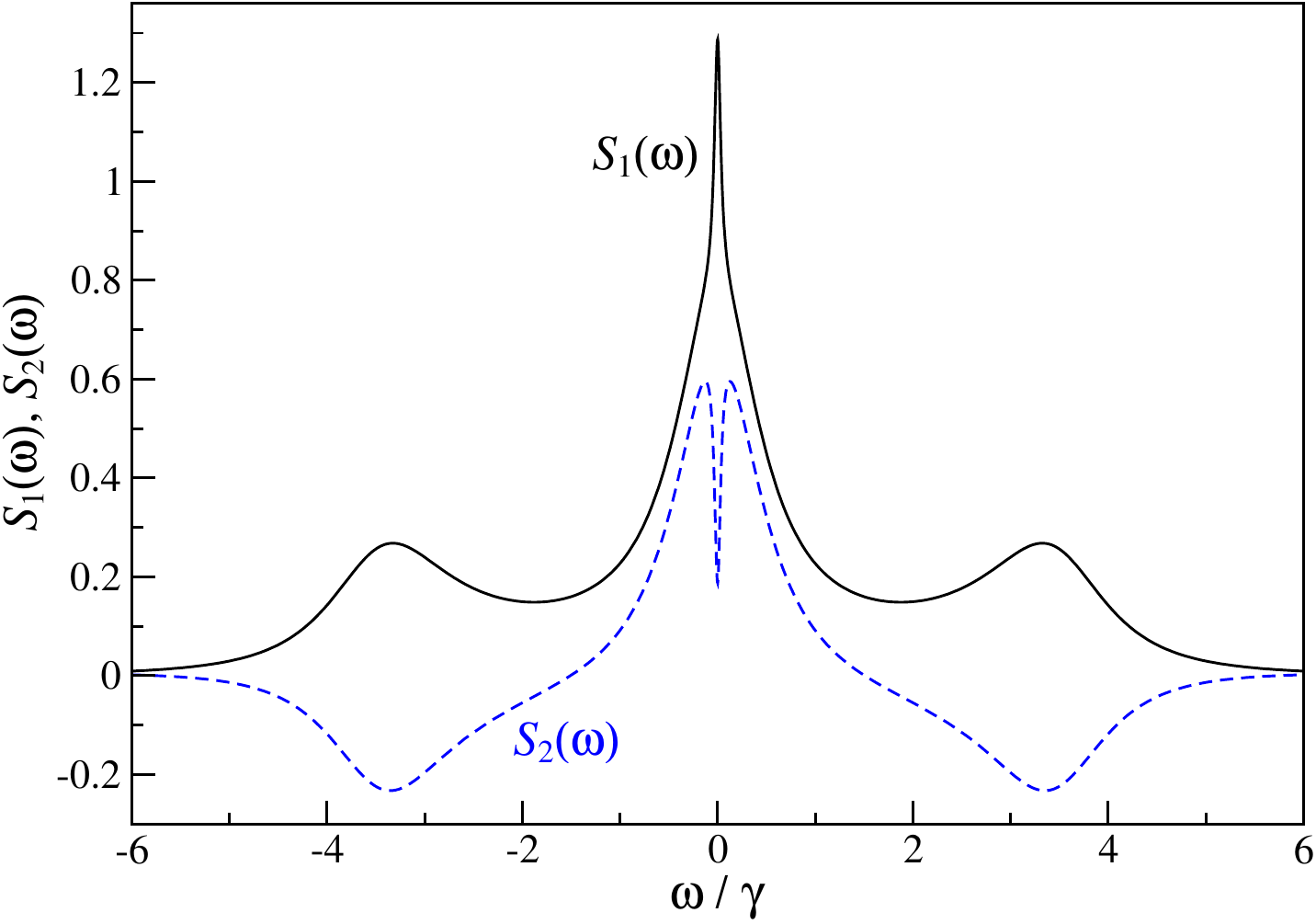}%
\caption{\label{fig:Specs1-2} 
(Color online) Spectra of the noise correlations 
$ \langle \Delta \sigma_+(0) \Delta \sigma_{+}(\tau) \rangle_{st}$ 
($S_1(\omega)$, solid line) and 
$\langle \Delta \sigma_+(0) \Delta \sigma_{-}(\tau) \rangle_{st}$ 
($S_2(\omega)$, dashed line for $\Omega=3.5 \gamma$,  
$\gamma_d=0.05 \gamma$, $\gamma_a=0.015 \gamma$, and $\eta=1$. }
\end{figure}

To further illustrate the difference among the spectra of quadratures, we 
plot in Fig. \ref{fig:Specs1-2} the spectra of the correlations 
$\langle \Delta \sigma_+(0) \Delta \sigma_{\mp}(\tau) \rangle_{st}$. For 
$S_{0}(\omega)$ the integrals are added, while for $S_{\pi/2}(\omega)$ 
they are subtracted. Thus, the sharp peak appears only in the latter. The 
addition or subtraction cancels spectral components. The spectrum 
(\ref{eq:SincDef}) contains only one of the integrals. 

\subsection{\label{sec:variance}Variances and integrated spectra} 
\begin{figure}[t]
\includegraphics[width=8.5cm]{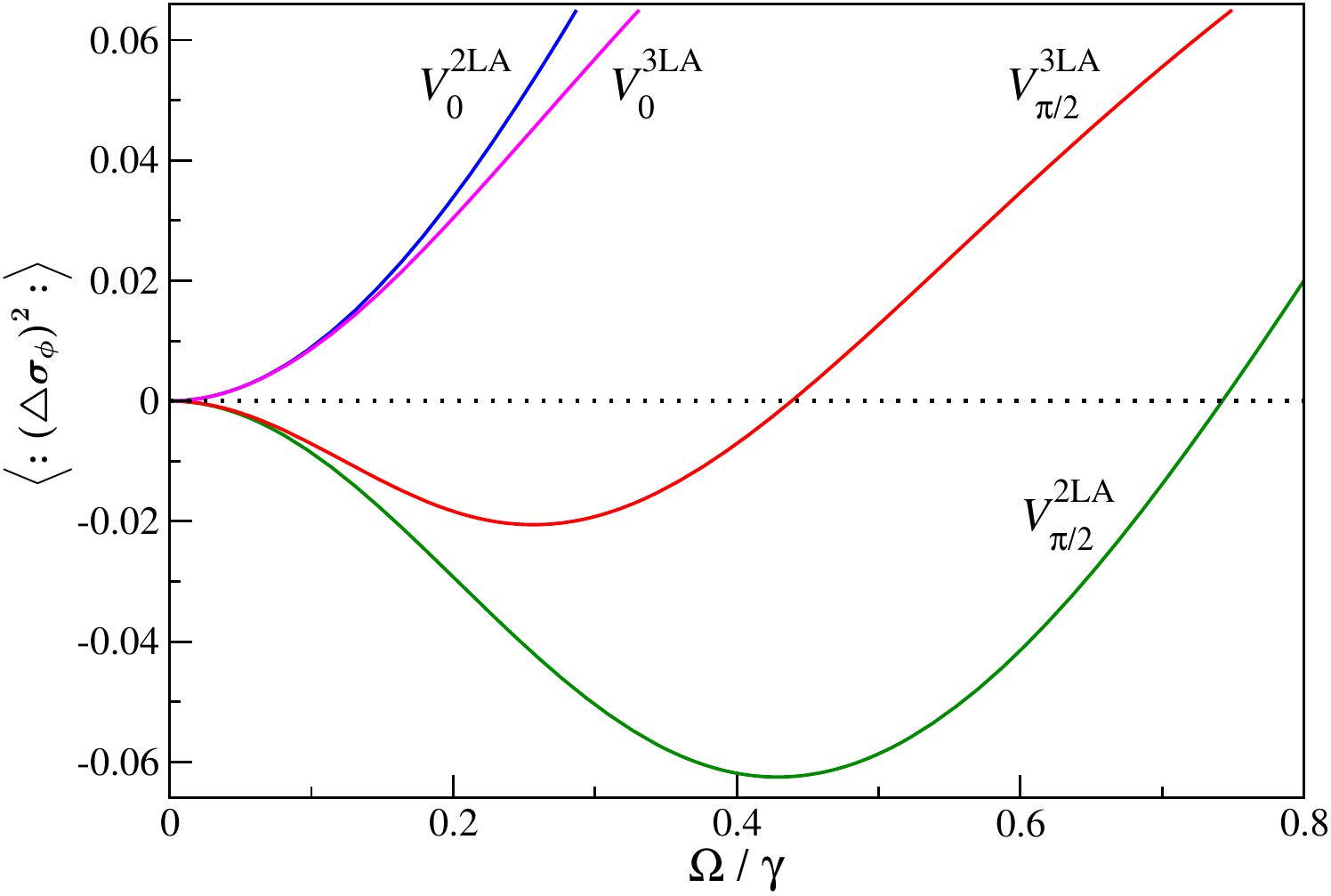}%
\caption{\label{fig:variance} 
(Color online) Variance of two and three-level atom resonance fluorescence 
for weak to moderately strong excitation, using Eqs. (\ref{eq:varphi}).  
 Additional parameters for the 3LA are $\gamma_d=0.05 \gamma$,  
$\gamma_a=0.015 \gamma$, and $\eta=1$. }
\end{figure}

An alternative approach to squeezing is the study of the variance or noise 
in a quadrature, 
\begin{eqnarray} 	\label{eq:variance}
V_{\phi} &=& \langle : (\Delta \sigma_{\phi})^2 : \rangle_{st}   
	=\mathrm{Re} \left[ e^{-i\phi}  
	\langle \Delta \sigma_+ \Delta \sigma_{\phi} \rangle_{st}  \right]   
\end{eqnarray}
or, equivalently, the integrated spectrum, related as $\int_{-\infty}^{\infty} 
S_{\phi}(\omega) d \omega =4\pi \gamma_+ \eta V_{\phi}$. A negative 
variance is a signature of squeezing in a quadrature. We have   
\begin{subequations}  	\label{eq:varphi} 
\begin{eqnarray} 	 
V_0 &=& 2 C_1  = \frac{Y^2/2}{ 1+Y^2 +(q/2)Y^2 }   \,, 	\\
%
V_{\pi/2}  &=& 2 ( C_2 -C_+ -C_- ) 	\nonumber \\ 
&=& \frac{Y^2/2}{ (1+Y^2) \left( 1+Y^2 +(q/2)Y^2 \right)^2}  \nonumber \\ 
&& \times \left[ Y^4 \left( 1+ \frac{q}{2} \right) +\frac{q}{2} Y^2 -1 \right] \,. 	
\end{eqnarray}
\end{subequations} 
We plot the variances in Fig. \ref{fig:variance}. $V_0$ is positive for any 
laser strength; there is no squeezing for $\phi=0$ but the total noise is 
smaller for the 3LA. For $V_{\pi/2}$ both the interval of the laser strength 
and amplitude for squeezing are notably reduced by the coupling to the 
long-lived state, and the Rabi frequency for the largest negative value is 
now very close to the saturating value, $\Omega=\gamma_+/4$, which 
we use for several spectra.  

The standard BHD technique depends on the finite detector efficiency 
$\eta$. This is a key obstacle to observe the weak squeezing of single-atom resonance fluorescence. Only very recently has the squeezing in the 
fluorescence of a single two-level quantum dot been observed 
\cite{SHJ+15} with homodyne correlation measurements 
\cite{Vogel91,Vogel95}, which are independent of the detector efficiency. 
However, the measured variance had to be extracted from complementary 
measurements with different phases. 

There is a subtle issue that also has to be addressed: Why is it that the 
quadrature variances are different? It seems natural to think that one 
features squeezing and the other does not. But, from the viewpoint of 
integrated spectra, one could expect this to be independent of the local 
oscillator phase. Thus, we reformulate the question: What spectrum could 
be integrated that gives the same value for both quadratures?

Conditional homodyne detection also solves the issue of finite detector 
efficiency, measuring an amplitude-intensity correlation, in this case 
without the need to extract the desired correlation from complementary 
measurements. CHD has been used to detect squeezing of a cavity QED 
source \cite{FOCC00}. Additionally, CHD gives an answer to the missing 
term in the integrated spectra. We devote the next two sections to a 
summary of CHD theory and its application to 3LA resonance fluorescence.

\section{\label{sec:chd-time}Conditional Homodyne Detection}  
Figure \ref{fig:chdsetup} illustrates the setup for amplitude-intensity 
correlation by CHD. Its theory was first presented in \cite{CCFO00}; its 
application to resonance fluorescence of a 2LA was given in 
\cite{hmcb10,CaGH15}, and its application to that of a $V$-type 
three-level atom was presented in \cite{MaCa08,CaGM14,XGJM15}. 
Hence, here we show only its basic features. A quadrature of the field, 
$E_{\phi}$, is measured in balanced homodyne detection conditioned on 
the direct detection of a photon (intensity, $I$) at detector $D_I$, i.e., 
$\langle I(0)E_{\phi}(\tau) \rangle_{st}$. Here, 
$E_{\phi} \propto \sqrt{\eta} \sigma_{\phi}$ and 
$I \propto \eta \sigma_+ \sigma_-$. Upon normalization, the 
dependence of the correlation on the detector efficiency $\eta$ is 
canceled. Then  
\begin{eqnarray} 	\label{eq:hDef} 
h_{\phi}(\tau) &=& 
	\frac{ \langle: \sigma_+(0) \sigma_-(0) \sigma_{\phi}(\tau) :\rangle_{st}}  
	 { \langle \sigma_+ \sigma_- \rangle_{st} \langle \sigma_{\phi} 
	\rangle_{st} } 	\,, 	
\end{eqnarray}
where it is assumed that the system is stationary, 
\begin{equation} 	\label{eq:dipoleQuad}
\sigma_{\phi}= 
  \frac{1}{2} \left(\sigma_- e^{i\phi} +\sigma_+ e^{-i\phi} \right) 
\end{equation}
is the dipole quadrature operator, $\phi$ is the phase between the strong 
local oscillator and the driving field, and we recall that $::$ indicates time and 
normal operator orderings. These orderings lead to different formulas for 
positive and negative time intervals, and in general, the correlations are 
asymmetric 
\cite{CCFO00,FOCC00,DeCC02,MaCa08,CaGM14,XGJM15,XuMo15}. 
However, in the present case the correlation is symmetric; thus we only 
use the expression for positive intervals: 
\begin{eqnarray} 	\label{eq:hDef2} 
h_{\phi}(\tau) &=& 
\frac{ \langle \sigma_+(0) \sigma_{\phi}(\tau) \sigma_-(0)  \rangle_{st}}  
 { \langle \sigma_+ \sigma_- \rangle_{st} \langle \sigma_{\phi} 
	\rangle_{st} } 	\,. 		
\end{eqnarray}

When the laser excites the atom on resonance, as is the case in this paper, 
the in-phase quadrature $\langle \sigma_{\phi=0}(t) \rangle$ vanishes at all 
times, and likewise 
$\langle \sigma_+(0) \sigma_{0}(\tau) \sigma_- (0)\rangle_{st}=0$. So, to 
obtain a finite measurement of this quadrature, it is necessary to add a 
coherent offset of amplitude $E_{\mathrm{off}}$ and phase $\phi=0$ to the 
dipole field before reaching the beam splitter \cite{hmcb10}. This procedure, 
however, hides the non-classical character of the fluorescence, showing a 
monotonously decaying correlation:  
\begin{eqnarray} 	\label{eq:h_0} 
h_{0}(\tau) &=&  1+ \frac{\alpha_{ee}}{\alpha_{ee} +E_{\mathrm{off}}^2 } 
	e^{-\gamma_+ \tau /2} 		 \,. 	
\end{eqnarray}

The $\phi=\pi/2$ quadrature is more interesting. Substituting 
Eqs. (\ref{eq:BlochSols}a), (\ref{eq:3rdordercorr}) and (\ref{eq:alphaee}) 
into Eq. (\ref{eq:hDef2}) we obtain
\begin{eqnarray} 	\label{eq:hhpi} 
h_{\pi/2}(\tau) &=& 1 +B_2 e^{\lambda_2 \tau} 
	-B_+ e^{\lambda_+ \tau}  -B_- e^{\lambda_- \tau}  	\,, 
\end{eqnarray}
where
\begin{subequations} 
\begin{eqnarray} 	\label{eq:B2pm} 
B_2 &=& q \frac{Y^2/2}{1+Y^2} 		\,, 	\\ 
B_{\pm} &=& \left( 1+ q \frac{Y^2/2}{1+Y^2} \right)
 \left( \frac{1}{2} \pm  \frac{1-2Y^2} {8 \delta/\gamma_+} \right)	\,. 
\end{eqnarray}
\end{subequations} 
The coupling to the metastable level $|a \rangle$ has visible consequences 
for both short and long times, making the CHD correlation amplitude larger 
than is the case for a 2LA, through the factor $q=\gamma_d/\gamma_a$. 
This excess amplitude decays slowly towards the unit value, which signals 
the decorrelation for long $\tau$, best noticed for large $\Omega$. 

The CHD correlation can be written in terms of correlations of fluctuation 
operators, as is the case with the full incoherent and squeezing spectra. 
Splitting the dipole operators into a mean plus fluctuations, Eq. 
(\ref{eq:dipoleSplit}), $h_{\phi}(\tau)$ is decomposed into a constant 
term plus two two-time correlations, one of second order and one of 
third order in the dipole fluctuation operators, 
\begin{subequations} 
\begin{eqnarray} 	\label{eq:hsplit} 
h_{\phi}(\tau) =1 +h_{\phi}^{(2)}(\tau) +h_{\phi}^{(3)}(\tau)  \,,
\end{eqnarray}
where 
\begin{eqnarray} 	\label{eq:h2} 
h_{\phi}^{(2)}(\tau) &=& \frac{ 2\mathrm{Re} \left[ 
	\langle \sigma_{-} \rangle_{st} \langle \Delta \sigma_+(0) 
	\Delta \sigma_{\phi}(\tau) \rangle_{st} \right] }
  {\langle \sigma_{\phi} \rangle_{st} \langle \sigma_+ \sigma_- \rangle_{st} } \,,  
\end{eqnarray}
\begin{equation} 	\label{eq:h3} 
h_{\phi}^{(3)}(\tau) =\frac{ \langle \Delta \sigma_+(0) 
	 \Delta \sigma_{\phi}(\tau) \Delta \sigma_- (0) 
	\rangle_{st} }{\langle \sigma_{\phi} \rangle_{st} 
	\langle \sigma_+ \sigma_- \rangle_{st} } \,.   	
\end{equation}
\end{subequations}
The splitting is not done by the measurement scheme, but it can be 
calculated to provide valuable information about the system's fluctuations. 
\begin{figure}[t]
\includegraphics[width=8.5cm]{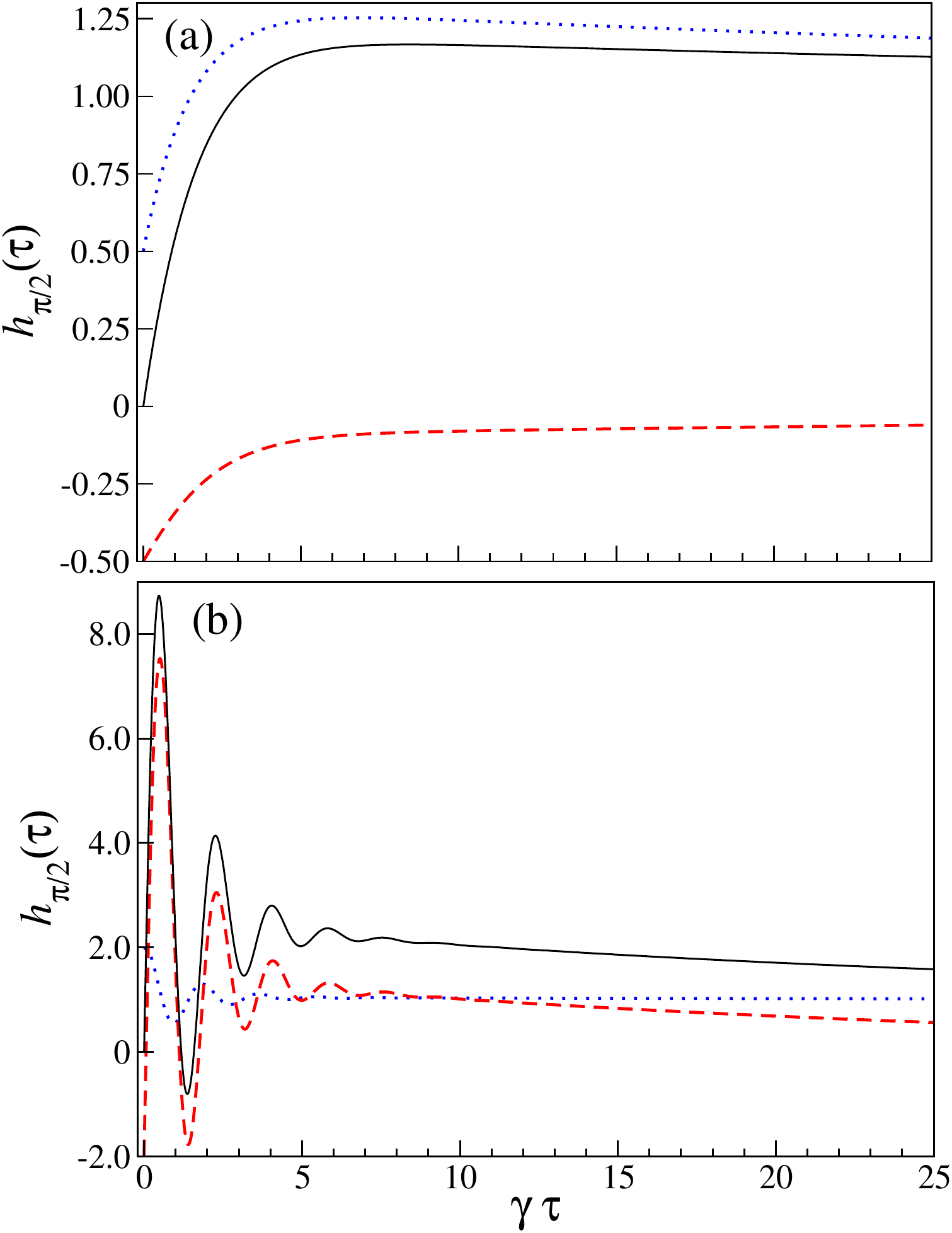}%
\caption{\label{fig:h_tau} 
(Color online) Amplitude-intensity correlation $h_{\pi/2}(\tau)$ (solid 
black line)  and its parts $1+h_{\pi/2}^{(2)}(\tau)$ (dotted blue line) and 
$h_{\pi/2}^{(3)}(\tau)$ 
(dashed red line) for (a) $\Omega=\gamma_+/4=0.2625 \gamma$ and  
(b) $\Omega=3.5 \gamma$. The other parameters are 
$\gamma_d=0.05 \gamma$ and $\gamma_a=0.015 \gamma$. Only the 
analytical results are plotted. }
\end{figure}

For $\phi=0$, due to the need to add an offset, we are left with 
Eq. (\ref{eq:h_0}). For $\phi=\pi/2$ we obtain the approximate expression: 
\begin{subequations} 
\begin{eqnarray} 	
h_{\pi/2}^{(2)}(\tau) &=& \frac{2}{\alpha_{ee}} \left[ C_2 e^{\lambda_2 \tau} 
	-C_+ e^{\lambda_+ \tau} -C_- e^{\lambda_- \tau} \right] 	\,, 
  \label{eq:h2ap} \\ 
h_{\pi/2}^{(3)}(\tau) &=& D_2 e^{\lambda_2 \tau} 
	+D_+ e^{\lambda_+ \tau} +D_- e^{\lambda_- \tau} \,,   \label{eq:h3ap}
\end{eqnarray}
where 
\begin{eqnarray} 	\label{eq:Dcoefs} 
D_{2} &=& B_2 -\frac{2 C_2}{\alpha_{ee} }	\,, 	\qquad
D_{\pm} = \frac{2 C_{\pm}}{\alpha_{ee}} -B_{\pm}	\,,
\end{eqnarray}
\end{subequations}
which are too cumbersome to be reproduced in full here. In Fig. 
\ref{fig:h_tau} we plot the analytical results, Eq. (\ref{eq:hhpi}) and its 
partial results $1+h_{\pi/2}^{(2)}(\tau)$ and $h_{\pi/2}^{(3)}(\tau)$, which 
differ very little from the exact ones. 

The vanishing of Eq. (\ref{eq:hsplit}) at $\tau=0$ has the same origin as 
the  antibunching in the intensity correlations: when the atom is in the 
ground state upon a photon emission, both the dipole field and the 
intensity are zero, and they build up again when the atom reabsorbs 
light. For $\phi=0$ the effect is not seen due to the additional offset.  So      
\begin{subequations}
\begin{eqnarray} 	\label{eq:h2_pi/2_0tau}
h_{\pi/2}^{(2)}(0) &=& \frac{\alpha_{ee} -2|\alpha_{+}|^2}{\alpha_{ee}} 	
	\nonumber \\
	&=& \frac{ Y^2 +(q/2)Y^2 -1 }{ 1+Y^2 +(q/2)Y^2 }	\,, 	
\end{eqnarray} 
and
\begin{eqnarray} 	
h_{\pi/2}^{(3)}(0) &=& \frac{2( |\alpha_{+}|^2 -\alpha_{ee})}{\alpha_{ee}}  
	= - 2(2+q) \alpha_{ee} 		\nonumber 	\\	
&=& -\frac{ (2+q) Y^2 }{1+Y^2 +(q/2)Y^2 } 	\,, 
	\label{eq:h3_pi/2_0tau} 
\end{eqnarray} 
\end{subequations}
that is, the initial size of the correlation is proportional to the mean 
population in the excited state. For $\Omega \gg \gamma$, the third-order 
correlation has its largest (negative) initial value $h_{\pi/2}^{(3)}(0) \to -2$.

The third-order term signals the deviation from Gaussian fluctuations as a 
consequence of the nonlinearity of the resonance fluorescence process 
for increasing laser intensity \cite{hmcb10,CaGH15}.  As perhaps best 
noticed in the spectral domain, it is the enhanced sensitivity to nonlinearity 
that makes CHD stand out over BHD and the spectrum of squeezing. 
We illustrate this in the next section.

\section{\label{sec:chd-spec}Quadrature Spectra from CHD} 
The spectrum measured from the amplitude-intensity correlation is given by 
\begin{eqnarray} 		 \label{eq:S-chd} 
\mathcal{S}_{\phi}(\omega) &=& 4 \gamma_+  \alpha_{ee} 
	\int_{0}^{\infty}  d\tau \cos{\omega \tau} 
	\left[ h_{\phi}(\tau) -1 \right]  \,.   
\end{eqnarray}
The factor $4 \gamma_+  \alpha_{ee}$ is the photon flux into the CHD 
setup. For $\phi=0$ we replace it by 
$4 \gamma_+  (\alpha_{ee} +E_{\mathrm{off}}^2)$.   
Following the splitting of $h_{\phi}(\tau)$, Eq. (\ref{eq:hsplit}), 
the spectra of second- and third-order dipole fluctuations are, respectively,  
\begin{subequations} 
\begin{eqnarray} 	 \label{eq:S2} 
\mathcal{S}_{\phi}^{(2)}(\omega) &=& 4\gamma_+ \alpha_{ee} 
	\int_{0}^{\infty}  d\tau \cos{\omega \tau} \,h_{\phi}^{(2)}(\tau)   
	\label{eq:S2} 	\,,	 \\ 
\mathcal{S}_{\phi}^{(3)}(\omega) &=& 4\gamma_+ \alpha_{ee}  
\int_{0}^{\infty} d\tau \cos{\omega \tau} \,h_{\phi}^{(3)}(\tau) 	\,,
	\label{eq:S3} 
\end{eqnarray}
\end{subequations} 

Using Eqs. (\ref{eq:h_0}) and (\ref{eq:hhpi}) we obtain the 
approximate analytical spectra. For $\phi=0$, we have
\begin{eqnarray} 		 \label{eq:S-chd-0} 
\mathcal{S}_{0}(\omega) &=& -4\gamma_+ \alpha_{ee}  
	 \frac{\lambda_1}{\omega^2 +\lambda_1^2} 	  \,,  
\end{eqnarray}
which is independent of the offset. The spectrum of this quadrature is a 
simple Lorentzian of width $\gamma_+/2$.  For $\phi=\pi/2$ we have 
\begin{eqnarray} 		 \label{eq:S-chd-hpi} 
\mathcal{S}_{\pi/2}(\omega) &=& 4\gamma_+ \alpha_{ee} 
	\left[ B_+ 	\frac{\lambda_+}{\omega^2 +\lambda_+^2} 
	+B_- \frac{\lambda_-}{\omega^2 +\lambda_-^2}
	\right.	\nonumber \\
&& \left.  - B_2	 \frac{\lambda_2}{\omega^2 +\lambda_2^2}    \right]  \,.   
\end{eqnarray}
The second-order spectra are
\begin{subequations} 
\begin{eqnarray} 	 
\mathcal{S}_{0}^{(2)}(\omega) &=& \mathcal{S}_{0}(\omega) 	\,,	\\
\mathcal{S}_{\pi/2}^{(2)}(\omega) &=& 8 \gamma_+
	\left[ C_+ 	\frac{\lambda_+}{\omega^2 +\lambda_+^2} 	
	+C_-	 \frac{\lambda_-}{\omega^2 +\lambda_-^2}  
	\right. 	\nonumber \\ 
&& \left.	-C_2 \frac{\lambda_2}{\omega^2 +\lambda_2^2}  \right]  \,.   
	\label{eq:S2app} 	 
\end{eqnarray}
\end{subequations} 
These are just the spectra of squeezing, Eqs. (\ref{eq:SqSpec_0}) and (\ref{eq:SqSpec_hpi}), without the detector efficiency factor.  
The third-order spectra are  
\begin{subequations} 
\begin{eqnarray}
\mathcal{S}_{\phi}^{(3)}(\omega) 
 &=& \mathcal{S}_{\phi}(\omega) -\mathcal{S}_{\phi}^{(2)}(\omega)  \\ 
\mathcal{S}_{0}^{(3)}(\omega) &=& 0 \,, 	\\ 
\mathcal{S}_{\pi/2}^{(3)}(\omega) 
&=& -4 \gamma_+ \alpha_{ee} \sum_{k=2,+,-} D_k 
	\frac{\lambda_k}{\omega^2 +\lambda_k^2} 	\,,	
\end{eqnarray} 
\end{subequations} 
where $D_k$ are given by Eq. (\ref{eq:Dcoefs}) and $\lambda_k$ are the 
eigenvalues Eq. (\ref{eq:eigenvalues}). 

Originally, CHD was conceived to overcome the issue of imperfect 
detection and thus be able to measure squeezing of weak light sources 
\cite{CCFO00,FOCC00}. In the weak-field limit the spectrum of the 
amplitude-intensity correlation approaches the spectrum of squeezing if 
third-order fluctuations can be neglected, i.e., 
\begin{eqnarray} 
S_{\phi}(\omega) = \eta \mathcal{S}_{\phi}^{(2)}(\omega) 
	\approx  \eta \mathcal{S}_{\phi}(\omega)	\,.
\end{eqnarray}
For the third-order spectrum, the sharp peak is about half the second-order 
one, of size $\sim Y^4$, while the other terms go also as $Y^4$, and the 
other second-order terms go as $Y^2$. However, for not-so-weak fields, 
we find strong signatures of third-order fluctuations in the spectra. 

\begin{widetext}
\begin{figure}[t]
\includegraphics[width=17.5cm]{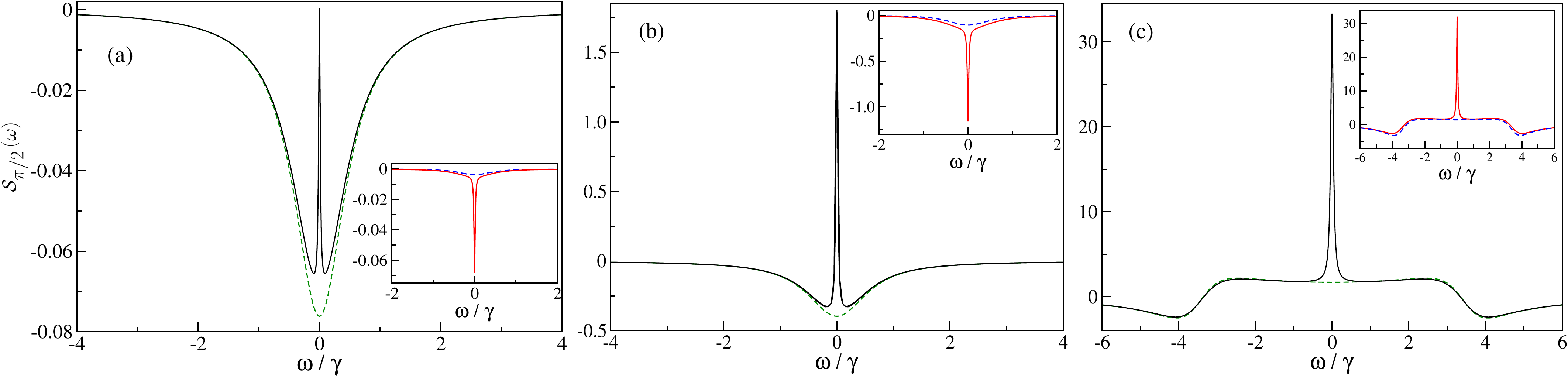}%
\caption{\label{fig:Schd} 
(Color online) Spectra of the amplitude-intensity correlation for $\phi=\pi/2$ 
of the three-level atom (solid black line) and two-level atom (dashed green 
line): (a) weak field $\Omega=0.1 \gamma$,  (b) moderate field
$\Omega=\gamma_+/4=0.2625 \gamma$, and (c) strong field 
$\Omega=3.5 \gamma$. The insets show the third-order spectra of the 
3LA (solid red line) and 2LA (dashed blue line). The second-order spectra 
are those of the spectrum of squeezing (Fig. \ref{fig:SqueezSpec}). Only 
the analytical results are plotted.}
\end{figure}
\end{widetext}

In Fig. \ref{fig:Schd} we plot the analytical results of the spectra of the 
amplitude-intensity correlation of the two- and three-level atom, Eq. 
(\ref{eq:S-chd-hpi}). The difference from the (omitted) exact results is very 
small. Note that for weak and saturating laser the sharp peak is smaller 
than in the spectra of squeezing, Fig. (\ref{fig:SqueezSpec}). This is 
because the third-order sharp peak is negative in this excitation regime, as 
seen in the insets. Moreover, in this excitation regime, the full third-order 
spectrum is negative [insets of Figs.~\ref{fig:Schd}(b) and \ref{fig:Schd}(c)], 
which adds to the negative squeezing peak of this quadrature 
\cite{hmcb10,CaGH15}. 

In the strong-excitation regime the third-order spectrum leads to striking 
deviations between the CHD and squeezing spectra and between the 
2LA and the 3LA. On the one hand, the sidebands become dispersive
\cite{hmcb10}. This comes out when $\lambda_{\pm}$ become complex, 
that is, for $\Omega > \gamma_+/4$, but it is only for strong enough 
excitation that the spectral components split. While the second-order 
peaks are Lorentzians, the third-order ones are dispersive and of 
comparable size for the 2LA \cite{hmcb10} or bigger for the 3LA. On the 
other hand, there are large deviations in the size of the spectra. The 
third-order spectrum is much bigger in the 3LA than in the 2LA, not only 
for the sharp peak. The third-order spectrum contributes most of the 
total CHD spectrum. 

The above effects can be explained as follows. The third-order correlation 
of fluctuation operators gives a measure of the atom-laser nonlinearity, 
which grows with increasing laser intensity, and the deviation from 
Gaussian fluctuations of the fluorescence. Also, it should be mentioned that 
the dipole fluctuations of the driven transition are enhanced due to the 
coupling to the long-lived state $| a \rangle$, which is populated by the 
increased number of spontaneous emission events from the excited state, 
Eq. (\ref{eq:alpha_aa}). An early study of this effect in the three-level 
configuration of this paper reported large deviations in the photon 
statistics from those of a 2LA \cite{MeSc90}.  

We recall that in CHD the second- and third-order components cannot 
be measured separately; both are merged in a single measured signal. 
CHD goes beyond the concept of squeezing when studying 
phase-dependent fluctuations. 

\subsection{\label{sec:variancechd}Integrated spectra} 
Finally, we calculate the integrated spectra of the CHD quadratures: 
\begin{subequations} 
\begin{eqnarray} 		 \label{eq:intS-chd} 
\int_{-\infty}^{\infty}  \mathcal{S}_{0}(\omega) d\omega 
	&=& 4\pi \gamma_+ \alpha_{ee} 	\,, 	\\
\int_{-\infty}^{\infty}  \mathcal{S}_{\pi/2}(\omega) d\omega 
	&=& -4\pi \gamma_+ \alpha_{ee}  	 \,. 	  
\end{eqnarray}
\end{subequations} 
That the magnitudes are equal means that the total emitted noise is 
independent of the quadrature. This is made possible by the third-order 
fluctuations, absent in the spectrum of squeezing, Sec. \ref{sec:variance}.  
This result is analogous to calculating the total incoherent emission by 
integrating the incoherent spectrum.  

\section{Conclusions}
We investigated ensemble-averaged phase-dependent fluctuations of the 
intermittent resonance fluorescence of a single three-level atom. We 
focused mainly on the spectrum of squeezing by balanced homodyne 
detection, and on the spectrum of the amplitude-intensity correlation of 
conditional homodyne detection. The shelving effect produces a sharp 
peak in the spectrum of the quadrature that features squeezing. Since this 
peak is positive, it acts to reduce the amount of squeezing observed in the 
weak- to moderate- (strong-) excitation regime. Since CHD is sensitive to 
third-order dipole fluctuations that grow with atom-laser nonlinearity, the 
spectra of BHD and CHD are very different for strong excitation. Additional 
insight is obtained by calculating the variances or integrated spectra of 
quadratures. In BHD the variances are different, while in CHD they are 
equal, a feature that deserves further study. 

We considered only the case of exact atom-laser resonance. This allowed 
us to obtain a very good approximate analytical solution of the master 
equation with a simple method, which we then used to construct analytical expressions for the various quantities of interest. Further insight into the 
incoherent spectrum and its link to the phase-dependent fluctuations could 
be established. Also, the on-resonance case allowed us to present the 
basic physical features in the most straightforward manner. 

Conditional homodyne detection, with its sensitivity to third-order field 
fluctuations, opens a new gate to study phase-dependent fluctuations 
beyond the realm of squeezing for highly nonlinear and non-Gaussian 
optical processes. On the other hand, the impressive advances in photon 
collection efficiencies by parabolic mirrors \cite{SoLe15,StSL10} could 
complement CHD for atomic resonance fluorescence, its squeezing and 
its quantum fluctuations in general.

\begin{acknowledgments}
H.M.C.-B thanks Prof. J. R\'ecamier for hospitality at ICF-UNAM. R.R.A 
thanks CONACYT for the scholarship No. 379732, and DGAPA-UNAM 
for support under project IN108413. 
\end{acknowledgments}

\appendix
\begin{widetext}
 
\section{\label{sec:approx}Approximate Solutions} 
The approximate expectation values of the atomic operators are
\begin{subequations} 	\label{eq:BlochSolsApp}
\begin{eqnarray} 	
\langle \sigma_{\mp} (t) \rangle &=& 
	\mp i \frac{ Y/ \sqrt{2} }{1+Y^2}  \left[ e^{\lambda_2 t} 
	 -e^{-3\gamma_+ t/4} \left( \cosh{\delta t}  
	+ \frac{3 \gamma_+}{4 \delta} \sinh{\delta t} \right) \right] 
  \mp i \sqrt{2} Y \frac{\gamma_+}{4 \delta} e^{-3\gamma_+ t/4} \sinh{\delta t}	
  +\alpha_{\mp} \left( 1- e^{\lambda_2 t} \right) \,,  	\nonumber \\   \\
\langle \sigma_{ee} (t) \rangle &=& 
	\frac{ Y^2/2 }{1+ Y^2}  \left[ e^{\lambda_2 t} 
	-e^{-3\gamma_+ t/4} \left( \cosh{\delta t}  
	+ \frac{3 \gamma_+}{4 \delta} \sinh{\delta t} \right)    \right] 
  	+\frac{Y^2/ 2}{1+Y^2 +(q/2)Y^2} \left( 1- e^{\lambda_2 t} \right) \,, 	\\ 
\langle \sigma_{gg} (t) \rangle &=& e^{\lambda_2 t}
	-\frac{ Y^2/2 }{1+ Y^2}   \left[ e^{\lambda_2 t} 
	-e^{-3\gamma_+ t/4} \left( \cosh{\delta t}  
	+ \frac{3 \gamma_+}{4 \delta} \sinh{\delta t} \right)   \right] 
  	+\frac{1+(Y^2/ 2)}{1+Y^2 +(q/2)Y^2} \left( 1- e^{\lambda_2 t} \right) \,. 	
\end{eqnarray}
\end{subequations} 

We make several assumptions to give our results simple, albeit long, 
expressions: First, we neglect a term $\gamma_d \Omega^2/2$ in the 
solutions in the Laplace space that reduce the problem to one similar to 
the 2LA case, with $\gamma$ replaced by $\gamma_+$. Eigenvalues 
$\lambda_{\pm}$ are thus identified. They give rise to the terms with the 
hyperbolic functions. Second, a constant term is multiplied by a factor 
$e^{\lambda_2 t}$. Finally, we add a term 
$\langle \sigma_{jk} \rangle_{st}(1-e^{\lambda_2 t})$. Recall that for the 
case of a 2LA $\lambda_2=0$ and 
$\langle \sigma_{ee} (t) \rangle +\langle \sigma_{gg} (t) \rangle =1$.  
These \textit{ad hoc} assumptions make the approximate solutions very  
close to the exact ones, as long as $\gamma_d$ and $\gamma_a$ are 
at least one order smaller than $\gamma$. 

Similarly, we obtain 
\begin{eqnarray} 	\label{eq:p_tauApp}
\langle \sigma_+(0)  \sigma_{\mp}(\tau) \rangle_{st} 
&=& \pm \frac{1}{2} \frac{ Y^2 }{ (1+Y^2 +(q/2)Y^2)^2 }
	+\frac{1}{4} \frac{Y^2}{ 1+Y^2 +(q/2)Y^2 } e^{\lambda_1 \tau} 
	\pm \frac{q}{4} \frac{Y^4}{(1+Y^2) \left( 1+Y^2 +(q/2)Y^2 \right)^2} 
		e^{\lambda_2 \tau} 
	\nonumber \\ 
&& \mp \frac{1}{4} \frac{Y^2 }{(1+Y^2) \left( 1+Y^2 +(q/2)Y^2 \right)} 
	e^{-3\gamma_+ \tau /4} \left[ (1-Y^2) \cosh{\delta \tau} 
	 \mp \frac{1-5Y^2}{4\delta/\gamma_+} \sinh{\delta \tau}  	\right] 	\,. 
\end{eqnarray}

\section{Equations of motion} 
We solve sets of linear equations of motion for the expectation values 
of the atomic operators and for two-time correlations. The equations 
and the formal solutions can be written as 
\begin{eqnarray} 	\label{eq:BlochEqsFluc} 
\frac{d}{dt} g(t) &=& \mathbf{M} g(t)  \,, 		\\ 
g(t) &=& e^{\mathbf{M} t} g(0) 	\,, 	
\end{eqnarray}
where $\mathbf{M}$ is the matrix (\ref{eq:matrixM}). In general, we solve 
these equations numerically. The initial conditions, however, are obtained 
exactly analytically, even off resonance. For instance, defining 
$\Delta \mathbf{s} \equiv  \left( \Delta \sigma_-, 
\Delta \sigma_+, \Delta \sigma_{ee}, \Delta \sigma_{gg}  \right)^T$, 
where $\Delta \sigma_{jk} =\sigma_{jk} -\alpha_{jk}$ and 
$\alpha_{jk} =\rho_{kj}^{st}$, $\alpha_+ =\rho_{ge}^{st}$, 
$\alpha_- =\rho_{eg}^{st}$, the initial conditions of the second- and 
third-order correlations of the fluctuation operators are 
\begin{eqnarray} 	\label{eq:corre2st}
\langle \Delta \sigma_+ \Delta \mathbf{s} \rangle_{st} 
= \left( \begin{array}{c} \alpha_{ee} -\alpha_+ \alpha_- \\  -\alpha_{+}^2 \\ 
	-\alpha_{+}\alpha_{ee} \\ 
	\alpha_{+} (1- \alpha_{gg}) \end{array} 	\right) 
= \frac{\Omega^2}{N^2} \left( \begin{array}{c} (2+q) \Omega^2  \\ 
	\gamma_+^2  \\ 	-i \gamma_+ \Omega 	\\
	i (1+q) \gamma_+ \Omega  	\end{array} \right) 	  \,,
\end{eqnarray} 
\begin{eqnarray} 	\label{eq:corre3st}
\langle \Delta \sigma_+ \Delta \mathbf{s} 
\Delta \sigma_- \rangle_{st} = \left( \begin{array}{c} 
  2\alpha_{-} (\alpha_+ \alpha_- -\alpha_{ee}) \\  
  2\alpha_{+} (\alpha_+ \alpha_- -\alpha_{ee}) \\ 
 \alpha_{ee} (2\alpha_+ \alpha_- -\alpha_{ee}) \\
  (\alpha_{gg} -1) (2\alpha_+ \alpha_- -\alpha_{ee}) \end{array}  \right) 
 = \frac{\Omega^4}{N^3} \left( \begin{array}{c} 
	i 2(2+q) \gamma_+ \Omega   	\\
	-i 2(2+q) \gamma_+ \Omega 	\\ 
	\gamma_+^2 - (2+q) \Omega^2  	\\
	(1+q) [ \gamma_+^2 -(2+q) \Omega^2  ] 	\end{array} \right) 	\,,
\end{eqnarray} 
respectively, where we used the steady-state values $\alpha_{jk}$ of 
Eqs. (\ref{eq:alphas}) and  $N= (2 +q)\Omega^2 +\gamma_+^2$. 

The numerical calculations of the spectra are more efficiently implemented 
using the formal solution of the correlations, 
$g(\tau)=e^{\mathbf{M} \tau} g(0)$, so the Fourier integral is formally 
solved as $(i \omega \mathbf{1-M})^{-1} g(0)$, where $\mathbf{1}$ is the 
$4 \times 4$ identity matrix. One is saved from potentially troublesome 
integrals where the upper limit is a long time of the order 
$\gamma_d^{-1}$.  
\end{widetext}




\begin{references} 
%
\bibitem{PlKn97} For a review see, e.g., M.~B. Plenio and P.~L. Knight, 
	Rev. Mod. Phys. \textbf{70}, 101 (1997). 
%
\bibitem{StHB09} F.~D. Stefani, J.~P. Hoogenboom, and E. Barkai, 
	Phys. Today \textbf{62}(2), 34 (2009).
%
\bibitem{HePl95} G.~C. Hegerfeldt and M.~B. Plenio, Phys. Rev. A 
	\textbf{52}, 3333 (1995). 
%
\bibitem{GaKK95} B.~M. Garraway, M.~S. Kim, and P.~L. Knight, 
	Opt. Commun. \textbf{117}, 560 (1995).
%
\bibitem{EvKe02} J. Evers and Ch. H. Keitel, Phys. Rev. A 
	\textbf{65}, 033813 (2002). 
%
\bibitem{BuTa00} V. B\"uhner and Chr. Tamm, Phys. Rev. A 
	\textbf{61}, 061801 (2000). 
%
\bibitem{HBLW97} J.~T. H\"offges, H.~W. Baldauf, W. Lange, and 
	H. Walther, J. Mod. Opt. \textbf{44}, 1999 (1997).  
%
\bibitem{WaZo91} D.~F. Walls and P. Zoller, Phys. Rev. Lett. 
	\textbf{47}, 709 (1981).  
	%
\bibitem{CoWZ84} M.~J. Collett, D.~F. Walls, P. Zoller, Opt. Commun. 
	\textbf{52}, 145 (1984).
%
\bibitem{SoLe15} M. Sonderman and G. Leuchs, in \textit{Engineering 
  the Atom-Photon Interaction}, edited by A. Predojevic and M.~W. Mitchell  
  (Springer, Heidelberg, 2015).
%
\bibitem{Vogel91} W. Vogel, Phys. Rev. Lett. \textbf{67}, 2450 (1991). 
%
\bibitem{Vogel95} W. Vogel, Phys. Rev. A \textbf{51}, 4160 (1995).
%
\bibitem{CCFO00} H.~J. Carmichael, H.~M. Castro-Beltran, G.~T. 
	Foster, and L.~A. Orozco, Phys. Rev. Lett. \textbf{85}, 1855 (2000).
%
\bibitem{FOCC00} G.~T. Foster, L.~A. Orozco, H.~M. Castro-Beltran, 
	and H.~J. Carmichael, Phys. Rev. Lett. \textbf{85}, 3149 (2000).  
%
\bibitem{HSL+06} H.-G. Hong, W. Seo, M. Lee, W. Choi, J.-H. Lee, and 
	K. An, Opt. Lett. \textbf{31}, 3182 (2006).
%
\bibitem{GSS+07} N.~B. Grosse, Th. Symul, M. Stobi\'nska, T.~C. Ralph, 
	and P.~K. Lam,  Phys. Rev. Lett. \textbf{98}, 153603 (2007).
%
\bibitem{KAD+09} L.~A. Krivitsky, U.~L. Andersen, R. Dong, A. Huck, 
 C. Wittmann, and G. Leuchs, Phys. Rev. A \textbf{79}, 033828 (2009).
%
\bibitem{KuVo-X} B. K\"uhn and W. Vogel, Phys. Rev. Lett. \textbf{116}, 
163603 (2016); arXiv:1511.01723.
%
\bibitem{Carm85} H.~J. Carmichael, Phys. Rev. Lett. \textbf{55}, 2790 (1985).
%
\bibitem{SHJ+15} C.~H.~H. Schulte, J. Hansom, A.~E. Jones, 
	C. Matthiesen, C. Le Gall, and M. Atat\"ure, Nature (London) 
	\textbf{525}, 222 (2015). 
%
\bibitem{GRS+09} S. Gerber, D. Rotter, L. Slodi\v{c}ka, J. Eschner, 
	H.~J. Carmichael, and R. Blatt, Phys. Rev. Lett. \textbf{102}, 
	183601 (2009).
%
\bibitem{hmcb10} H.~M. Castro-Beltran, Opt. Commun. \textbf{283}, 
	4680 (2010).
%
\bibitem{CaGH15} H.~M. Castro-Beltran, L. Gutierrez, and L. Horvath, 
	Appl. Math. Inf. Sci. \textbf{9}, 2849 (2015).
%
\bibitem{DeCC02} A. Denisov, H.~M. Castro-Beltran, and 
	H.~J. Carmichael, Phys. Rev. Lett. \textbf{88}, 243601 (2002).
%
\bibitem{MaCa08} E.~R. Marquina-Cruz and H.~M. Castro-Beltran,  
	Laser Phys. \textbf{18}, 157 (2008).
%
\bibitem{CaGM14} H.~M. Castro-Beltran, L. Gutierrez, and E.~R. 
	Marquina-Cruz, in \textit{Latin America Optics and Photonics}, 
	Cancun, Mexico, 2014, OSA Technical Digest 
	(Optical Society of America, Washington, D.C., 2014),  paper LM4A.38.   
%
\bibitem{XGJM15} Q. Xu, E. Greplova, B. Julsgaard, and K. M\o lmer, 
	Phys. Scripta  \textbf{90}, 128004 (2015). 
%
\bibitem{XuMo15} Q. Xu and K. M\o lmer, Phys. Rev. A 
	\textbf{92}, 033830 (2015).
%
\bibitem{Carm99} H.~J. Carmichael, \textit{Statistical Methods in 
  Quantum Optics 1: Master Equations and Fokker-Planck Equations} 
  (Springer, Berlin, 2002).
%
\bibitem{Mollow69} B.~R. Mollow, Phys. Rev. \textbf{188}, 1969 (1969). 
%
\bibitem{RiCa88} P.~R. Rice and H.~J. Carmichael, J. Opt. Soc. Am. B, 
	\textbf{5}, 1661 (1988). 
%
\bibitem{Carm87} H.~J. Carmichael, J. Opt. Soc. Am. B, \textbf{4}, 1588 
(1987). The ideal source field spectrum of squeezing is the one produced 
by the (atomic) source alone, neglecting the free field. Since the latter is 
assumed to be in the vacuum state, its omission is justified on the basis 
of using normal and time operator orderings. 
%
\bibitem{MeSc90} M. Merz and A. Schenzle, Appl. Phys. B 
	\textbf{50}, 115 (1990). 
%
\bibitem{StSL10} M. Stobi\'nska, M. Sonderman, and G. Leuchs, 
	Opt. Commun. \textbf{283}, 737 (2010).
%

\end{references}
\end{document}